\documentclass{ifacconf}

\usepackage{graphicx}      % include this line if your document contains figures
\usepackage{natbib}        % required for bibliography
\usepackage{amsmath} 	
\usepackage{amssymb} 	

\usepackage{algorithm}
\usepackage{algorithmicx}
\usepackage{algpseudocode}

\usepackage{tikz}
\usepackage{pgfplots}  
\pgfplotsset{compat=1.18}
\usepgfplotslibrary{groupplots}
\usetikzlibrary{patterns}

\usepackage[acronym]{glossaries}

\newtheorem{problem}{Problem}
\newtheorem{definition}{Definition}

\newtheorem{lemma}{Lemma}
\newtheorem{remark}{Remark}

\newacronym{GT}{GT}{Ground Truth}
\newacronym{LQG}{LQG}{linear--quadratic Gaussian}
\newacronym{LQ}{LQ}{linear--quadratic}
\newacronym{ISDG}{ISDG}{Inverse Stochastic Differential Game}
\newacronym{IRL}{IRL}{Inverse Reinforcement Learning}
\newacronym{IDG}{IDG}{Inverse Differential Game}

\newcommand{\qedsymbol}{$\hfill\square$}

\usepackage{fancyhdr}
\fancypagestyle{firstpage}{
  \fancyhf{}%
  \fancyhead[R]{© 2026 the authors. This work has been accepted to IFAC for publication under a Creative Commons Licence CC-BY-NC-ND.}
  
}
\begin{document}

\begin{frontmatter}

\title{Inverse Linear--Quadratic Gaussian Differential Games} 
% Title, preferably not more than 10 words.

\author[First]{Lucas Günther}
\author[First]{Felix Thömmes}
\author[First]{Karl Handwerker}
\author[First]{Balint Varga} 
\author[First]{Sören Hohmann}

\address[First]{Institute of Control Systems, Karlsruhe Institute of Technology, 76131 Karlsruhe, Germany (e-mail: lucas.guenther@kit.edu).}

\begin{abstract}% Abstract of 50--100 words
This paper presents a method for solving the Inverse Stochastic Differential Game (ISDG) problem in finite-horizon linear--quadratic Gaussian (LQG) differential games. The objective is to recover cost function parameters of all players, as well as noise scaling parameters of the stochastic system, consistent with observed trajectories. The proposed framework combines (i) estimation of the feedback strategies, (ii) identification of the cost function parameters via a novel reformulation of the coupled Riccati differential equations, and (iii) maximum likelihood estimation of the noise scaling parameters. Simulation results demonstrate that the approach recovers parameters, yielding trajectories that closely match the observed trajectories.
\end{abstract}


\end{frontmatter}

\thispagestyle{firstpage}
%===============================================================================

\section{Introduction}\label{sec:introduction}
Non-cooperative differential games, first introduced by \cite{Isa:65}, provide a mathematical framework for modeling multi-agent systems in various engineering applications (see, e.g., \cite{mylvaganam_differential_2017, Var:21, gu_differential_2008}). The solution of a differential game, 
%such as a Nash equilibrium, 
also referred to as the forward problem, has received considerable attention in both deterministic (see, e.g., \cite{BasOls:99,Eng:05}) and stochastic (see, e.g., \cite{Ham:98, BucLi:08}) systems.
While the forward problem presupposes known cost functions and derives the induced trajectories, the inverse problem aims to identify the underlying cost functions from observed, so-called \gls{GT} trajectories. To date, this has been mainly explored in single-agent settings, either within the framework of Inverse Optimal Control (see, e.g., \cite{JeaMas:18, MenZei:20}) or \gls{IRL} (see, e.g., \cite{AbbNg:04, Zie:08}).
A primary application of these inverse methods is the identification of the cost functions that explain observed human behavior.
In the context of motor control, this behavior is inherently stochastic due to variability and uncertainty in perception and decision-making, as described by \cite{TodJor:02}. 
Therefore, the inverse problem must estimate both the parameters of the cost function and the noise scaling terms governing the stochastic dynamics (\cite{Kar:24}).
Extending inverse stochastic methods to the multi-player setting remains largely unexplored, yet it is essential for applications such as human--machine interaction (\cite{Kil:24}). This motivates the development of \gls{ISDG} methods capable of identifying both the players’ cost function parameters and the system’s noise scaling terms from observed \gls{GT} trajectories, which we address in this paper.

%The remainder of this paper is structured as follows. In Section~\ref{sec:related_work}, we review related work and outline our contributions. In Section~\ref{sec:problem}, we state the \gls{ISDG} problem. In Section~\ref{sec:main_result}, we present our proposed method for solving the \gls{ISDG} problem. Section~\ref{sec:simulation} illustrates the approach with simulation results, and we conclude the paper in Section~\ref{sec:conclusion}.

\section{Related Work and Contributions}\label{sec:related_work}
\gls{IDG} methods can be categorized in direct and indirect approaches (\cite{Mol:22}). Direct methods minimize an error measure between the observed trajectories and the trajectories being optimal with respect to a current guess of the cost functions. This approach is computationally expensive, since it solves the differential game in each iteration to evaluate the error (\cite{Mol:22}). In contrast, indirect methods minimize the violation of the optimality conditions that are satisfied by the \gls{GT} trajectories and the unknown cost function parameters (\cite{Mol:22}). In this work, we focus on the indirect approach due to its better computational efficiency and scalability.

\subsection{Inverse Methods for Deterministic Games}

Indirect methods for \gls{IDG}s can be derived either from the Pontryagin Minimum Principle (PMP) (see, e.g., \cite{Mol:22,Rot:17}) or from the coupled Hamilton--Jacobi--Bellman (HJB) equations (see, e.g., \cite{LIAN2022110524}). 
In control theory, particular attention has been given to \gls{LQ} differential games, where linear system dynamics and quadratic cost functions lead to a reduction of the coupled HJB equations to a set of coupled Riccati differential equations (\cite{BasOls:99}, Corollary~6.5). 
Existing inverse methods based on the coupled HJB or Riccati formulation, however, are limited to the infinite-horizon case, where in \gls{LQ} differential games the coupled algebraic Riccati equations arise (see, e.g., \cite{Ing:19, MarCao:24}). In contrast, PMP-based approaches have been developed for the finite-horizon setting, but a formulation based on the coupled time-varying Riccati differential equations has not been addressed yet.

\subsection{Inverse Methods for Stochastic Games}

Inverse methods for stochastic games can be categorized according to the underlying formulation of the game. Methods developed for stochastic differential games address continuous-time dynamics, whereas those formulated for Markov or dynamic games operate in discrete-time and are typically solved via multi-agent \gls{IRL}.
Recent work on \gls{ISDG}s has investigated the \gls{LQG} setting, where system dynamics are linear, cost functions are quadratic, and the noise enters additively as Gaussian noise (\cite{Che:24a, Che:25}). 
In \gls{LQG} differential games, the resulting feedback strategies coincide with those of the corresponding deterministic \gls{LQ} differential game, which can be exploited to construct inverse methods. Both \cite{Che:24a, Che:25} consider two-player infinite-horizon \gls{LQG} differential games and rely on coupled algebraic Riccati equations. While \cite{Che:24a} study a general-sum setting and \cite{Che:25} focus on a zero-sum formulation, they both adopt an adaptive framework, in which selected cost function parameters and strategies are updated online based on observed trajectories. Their scope, however, is limited to identifying only a subset of the cost function parameters of a single player, rather than estimating the complete cost functions of all players. In addition, the stochastic components of the system, particularly the noise scaling parameters, are assumed to be known a priori.

Complementary to this, existing methods for recovering cost function parameters in discrete-time settings via multi-agent \gls{IRL} can be distinguished into two main approaches: early works based on the Bellman equation (\cite{Nat:10, Red:12}) and subsequent developments derived from the maximum entropy principle (\cite{Yu:19, Meh:23}). While these methods can be viewed as direct, their data-driven structure requires solving iterative optimization problems. Consequently, they are computationally demanding and data-intensive, primarily suited for settings with abundant data availability rather than real-time or data-limited applications, such as human–machine interaction. Moreover, none of these methods provide estimates of the system’s noise scaling parameters.

\subsection{Research Gap and Contributions}

Consequently, no method exists that recovers both the complete cost function parameters and the noise scaling parameters in a stochastic multi-player setting. The structure of \gls{LQG} differential games is particularly promising, as the cost function parameters can be obtained from the corresponding deterministic subproblem using coupled Riccati differential equations. However, for finite-horizon differential games, no inverse method has yet exploited this structure. To address this gap, this work makes two main contributions:
\begin{itemize}
    %\item A formal definition of the \gls{ISDG} problem, aiming to recover players' cost function and noise scaling parameters from observed state and control input trajectories.
    \item A novel \gls{IDG} method for finite-horizon \gls{LQ} differential games, formulated using coupled Riccati differential equations as equality constraints.
    \item A complete \gls{ISDG} framework for \gls{LQG} differential games that (i) estimates feedback strategies via least squares estimation, (ii) estimates cost function parameters using the novel \gls{IDG} method, and (iii) estimates noise scaling parameters via maximum likelihood estimation, thereby solving the \gls{ISDG} problem for the first time.
\end{itemize}

\section{Problem Formulation}\label{sec:problem}

Let the linear stochastic system dynamics be given by
\begin{equation}\label{eq:system}
    d\boldsymbol{x}(t) = \left(\boldsymbol{A}\boldsymbol{x}(t)+ \sum_{i=1}^N\boldsymbol{B}_i\boldsymbol{u}_i(t)\right)dt + \boldsymbol{L}
    d\boldsymbol{\omega}(t),
\end{equation}
where $\boldsymbol{x}(t) \in \mathbb{R}^n$ denotes the system state with initial condition $\boldsymbol{x}(t_0)=\boldsymbol{x}_0$, $\boldsymbol{u}_i(t) \in \mathbb{R}^{m_i}$ denotes the control input of player $i \in \{1, \dots,N\} =: \mathcal{I} \subset\mathbb{N}_+$, and $d\boldsymbol{\omega}(t) \in \mathbb{R}^n$ is the increment of an $n$-dimensional standard Wiener process satisfying $\mathbb{E}\left[d\boldsymbol{\omega}(t)\right] = \boldsymbol{0}$ and $\mathbb{E}\left[d\boldsymbol{\omega}(t)\left(d\boldsymbol{\omega}(t)\right)^\top\right] = \boldsymbol{I}_ndt$. The system matrices are given by $\boldsymbol{A} \in\mathbb{R}^{n\times n}, \boldsymbol{B}_i\in\mathbb{R}^{n\times m_i},\, \forall i \in \mathcal{I}$, and $\boldsymbol{L}\in\mathbb{R}^{n\times n}$, where $\boldsymbol{L}=\operatorname{diag}(l_1, \dots, l_n)$ with $l_s>0,\, \forall s \in \{1,\dots,n\}$, is referred to as the noise scaling matrix, so that the noise term has covariance $\mathbb{E}[(\boldsymbol{L}d\boldsymbol{\omega}(t))(\boldsymbol{L}d\boldsymbol{\omega}(t))^\top]=\boldsymbol{L}\boldsymbol{L}^\top dt$.

We consider players selecting their controls using linear feedback strategies of the form
\begin{equation}\label{eq:feedback_strategy}
    \boldsymbol{u}_i(t) =  - \boldsymbol{K}_i(t)\boldsymbol{x}(t),\, \forall i \in \mathcal{I},
\end{equation}
and restrict $\boldsymbol{K}_i(t)\in \mathbb{R}^{m_i \times n}$ such that the $N$-tuple of feedback strategies $\boldsymbol{K}(t) := \left(\boldsymbol{K}_1(t), \dots,\boldsymbol{K}_N(t)\right)$ stabilizes the closed-loop dynamics of \eqref{eq:system}. We further consider non-cooperative \gls{LQG} differential games, where each player $i$ aims to minimize an individual cost function
\begin{equation}\label{eq:cost_function}
\begin{aligned}
    J_i &= \mathbb{E}\Bigg[\boldsymbol{x}(t_N)^\top\boldsymbol{Q}_{i,t_N}\boldsymbol{x}(t_N)\\
    &+\int_{t_0}^{t_N}\boldsymbol{x}(t)^\top\boldsymbol{Q}_i\boldsymbol{x}(t)+ \sum_{j=1}^N \boldsymbol{u}_j(t)^\top\boldsymbol{R}_{ij}\boldsymbol{u}_j(t)\,dt\Bigg],
\end{aligned}
\end{equation}
with respect to \eqref{eq:system}, where $\boldsymbol{Q}_i, \boldsymbol{R}_{ij}, \boldsymbol{Q}_{i,t_N}$ are the symmetric weighting matrices for all $i \in \mathcal{I}$, with $\boldsymbol{Q}_i \succ \boldsymbol{0}$, $\boldsymbol{R}_{ii} \succ \boldsymbol{0}$, $\boldsymbol{R}_{ij} \succeq \boldsymbol{0},\, \forall j \in \mathcal{I} \setminus \{i\}$, $\boldsymbol{Q}_{i,t_N} \succeq \boldsymbol{0}$.

We write $J_i(\boldsymbol{x}_0,\boldsymbol{K}_i(t), \boldsymbol{K}_{\neg i}(t))$, where $\boldsymbol{K}_{\neg i}(t) := ( \boldsymbol{K}_j(t) \mid  j  \in  \mathcal{I}\setminus\{i\})$, as a function of feedback strategies and the initial state, since these generate the state and control trajectories $\boldsymbol{x}(t)$ and $\boldsymbol{u}_i(t),\,\forall i \in \mathcal{I}$, through \eqref{eq:system} and \eqref{eq:feedback_strategy}. We assume that the players' strategies result in a feedback Nash equilibrium, defined as follows.
\begin{definition}[\cite{BasOls:99}, Definition~3.12]\label{def:nash_equilibrium}
    An $N$-tuple $\boldsymbol{K}^*(t)$ is called a feedback Nash equilibrium if
    \begin{equation}\label{eq:feedback_nash_equilibrium}
    J_i(\boldsymbol{x}_0,\boldsymbol{K}_i^*(t), \boldsymbol{K}_{\neg i}^*(t)) \leq J_i(\boldsymbol{x}_0,\boldsymbol{K}_i(t), \boldsymbol{K}_{\neg i}^*(t))
    \end{equation}
    holds for all $i \in \mathcal{I}$ and all $\boldsymbol{x}_0\in \mathbb{R}^n$.
\end{definition}
%As motivated in Section~\ref{sec:related_work}, we now formally state the \gls{ISDG} problem for the \gls{LQG} setting.
\begin{problem}\label{prob:isdg}
    Let the $N$ players act in a feedback Nash equilibrium according to Definition~\ref{def:nash_equilibrium}, and let $D \in \mathbb{N}_+$ corresponding demonstrations, i.e., \gls{GT} trajectories $\{\boldsymbol{x}^{*(d)}(t_0 \to t_N)\}$ and $\{\boldsymbol{u}_i^{*(d)}(t_0 \to t_N)\},\,\forall i \in \mathcal{I}$, be given. Furthermore, let the system matrices $\boldsymbol{A}$ and $\boldsymbol{B}_i,\, \forall i \in \mathcal{I}$, be given. Find at least one combination of parameters $\hat{\boldsymbol{Q}}_i,\,\hat{\boldsymbol{R}}_{ij}, \hat{\boldsymbol{Q}}_{i,t_N}, \,\forall i,j \in \mathcal{I}$, and $\hat{\boldsymbol{L}}$ such that $\{\boldsymbol{x}^{*}(t_0 \to t_N)\} = \{\hat{\boldsymbol{x}}^*(t_0 \to t_N)\}$ and $\{\boldsymbol{u}_i^{*}(t_0 \to t_N)\} = \{\hat{\boldsymbol{u}}_i^{*}(t_0 \to t_N)\},\,\forall i \in \mathcal{I}$, where $\{\hat{\boldsymbol{x}}^*(t_0 \to t_N)\}$ and $\{\hat{\boldsymbol{u}}^*_i(t_0 \to t_N)\}$ correspond to a feedback Nash equilibrium resulting from $\hat{\boldsymbol{Q}}_i,\,\hat{\boldsymbol{R}}_{ij},\,\hat{\boldsymbol{Q}}_{i,t_N}, \,\forall i,j \in \mathcal{I}$, and $\hat{\boldsymbol{L}}$.
\end{problem}
\section{Inverse Linear-Quadratic Gaussian Differential Games}\label{sec:main_result}
To solve Problem~\ref{prob:isdg}, in this section, we introduce a novel inverse method for \gls{LQG} differential games. The proposed approach consists of three main steps. First, the $D$ \gls{GT} trajectories are used to identify the feedback Nash strategies $\boldsymbol{K}_i^*(t)$ of each player $i$ through least squares estimation. Second, the identified feedback Nash strategies $\boldsymbol{K}^*(t)$ are substituted into the coupled Riccati differential equations. This allows the coupled Riccati differential equations to be reformulated as $N$ decoupled differential equations that must hold for all $t \in [t_0;t_N]$. Based on this reformulation, we construct for each player $i$ a linear system of equations that can be solved to recover the unknown cost function parameters $\boldsymbol{Q}_i,\,\boldsymbol{R}_{ij},\,\boldsymbol{Q}_{i,t_N}, \,\forall i,j \in \mathcal{I}$. Finally, the noise scaling matrix $\boldsymbol{L}$ is estimated via maximum likelihood optimization using the discretized system dynamics and the \gls{GT} trajectories. 
%The section concludes with a summary of the complete \gls{ISDG} framework for \gls{LQG} differential games, integrating all three components into an estimation procedure.
\subsection{Identification of Feedback Nash Strategies}
Given the $D$ \gls{GT} trajectories, for each $t \in [t_0;t_N]$ we define
\begin{equation}
    \begin{aligned}
        \boldsymbol{X}(t) &= \left[\,\boldsymbol{x}^{*(1)}(t)\;\; \boldsymbol{x}^{*(2)}(t)\;\; \cdots\;\; \boldsymbol{x}^{*(D)}(t)\,\right]\in\mathbb{R}^{n\times D}\\
    \boldsymbol{U}_i(t) &= \left[\,\boldsymbol{u}_i^{*(1)}(t)\;\; \boldsymbol{u}_i^{*(2)}(t)\;\; \cdots\;\; \boldsymbol{u}_i^{*(D)}(t)\,\right]\in\mathbb{R}^{m_i\times D},
    \end{aligned}
\end{equation}
so that the linear feedback strategies at time $t$ read
\begin{equation}\label{eq:estimated_feedback_matrix_equaiton}
    \boldsymbol{U}_i(t) = -\boldsymbol{K}^*_i(t)\,\boldsymbol{X}(t).
\end{equation}
\begin{assum}\label{assum:percistence_excitation}
Let there exist a constant $\alpha>0$ such that $\boldsymbol{X}(t)\boldsymbol{X}(t)^\top \succeq \alpha I_n$ for all $t\in[t_0,t_N]$, ensuring persistence of excitation.
\end{assum}
    
\begin{lemma}\label{lem:iedentification_strategies}
    Let Assumption~\ref{assum:percistence_excitation} hold. Then the least squares estimator
    \begin{equation}\label{eq:exact_ls}
   \hat{\boldsymbol{K}}_i(t) = -\boldsymbol{U}_i(t)\,\boldsymbol{X}(t)^\top\big(\boldsymbol{X}(t)\boldsymbol{X}(t)^\top\big)^{-1}
\end{equation}
recovers the feedback strategy exactly, i.e., \(\hat{\boldsymbol{K}}_i(t)=\boldsymbol{K}_i^*(t)\).
\end{lemma}
\begin{pf}
    Since the \gls{GT} trajectories correspond to a feedback Nash equilibrium, we have \eqref{eq:estimated_feedback_matrix_equaiton}. By Assumption~\ref{assum:percistence_excitation} the matrix $\left(\boldsymbol{X}(t)\boldsymbol{X}(t)^\top\right)$ is positive definite and thus invertible. Substituting into \eqref{eq:exact_ls} yields
    \begin{equation}
    \hat{\boldsymbol{K}}_i(t) = \boldsymbol{K}^*_i(t)\,\boldsymbol{X}(t)\,\boldsymbol{X}(t)^\top\big(\boldsymbol{X}(t)\boldsymbol{X}(t)^\top\big)^{-1}= \boldsymbol{K}^*_i(t),
\end{equation}
which proves exact recovery, following standard least squares consistency (\cite{newey_large_nodate}, Section~2.2).\qedsymbol
\end{pf}
\subsection{Identification of Cost Function Parameters}
\begin{lemma}[\cite{BasOls:99}, Corollary~6.5]\label{lem:necessary_sufficient}\hfill
    Let there exist an $N$-tuple of symmetric matrices $\boldsymbol{P}_i(t) \in \mathbb{R}^{n\times n}$, $\forall i\in \mathcal{I}$, satisfying the $N$ coupled matrix Riccati differential equations
\begin{equation}\label{eq:coupled_riccati}
\begin{aligned}
   \dot{\boldsymbol{P}}_i(t) &= - \boldsymbol{P}_i(t) \boldsymbol{F}(t) -\boldsymbol{F}(t)^\top\boldsymbol{P}_i(t) -\boldsymbol{Q}_i\\
   &- \sum_{j=1}^N \boldsymbol{P}_j(t) \boldsymbol{B}_j \boldsymbol{R}_{jj}^{-1}\boldsymbol{R}_{ij}\boldsymbol{R}_{jj}^{-1}\boldsymbol{B}_j^\top\boldsymbol{P}_j(t),   
\end{aligned}
\end{equation}
where $\boldsymbol{P}_i(t_N)=\boldsymbol{Q}_{i,t_N}$ and
\begin{equation}
    \boldsymbol{F}(t)=\boldsymbol{A}-\sum_{i=1}^N\boldsymbol{B}_i\boldsymbol{K}_i(t) \in \mathbb{R}^{n\times n},
\end{equation}
such that the closed-loop dynamics of \eqref{eq:system} is stabilized. Furthermore, let $\boldsymbol{K}_i^*(t)$ be defined as
\begin{equation}\label{eq:relation_K_P}
    \boldsymbol{K}_i^*(t)=\boldsymbol{R}_{ii}^{-1}\boldsymbol{B}_i^\top\boldsymbol{P}_i(t).
\end{equation}
Then $\boldsymbol{K}^*(t)$ is a feedback Nash equilibrium as in Definition~\ref{def:nash_equilibrium}. Conversely, if $\boldsymbol{K}^*(t)$ is a feedback Nash equilibrium, then \eqref{eq:coupled_riccati} has a stabilizing solution.
\end{lemma}
\begin{pf}
    See (\cite{BasOls:99}, Corollary~6.5).\qedsymbol
\end{pf}
Since Lemma~\ref{lem:necessary_sufficient} provides a necessary and sufficient condition for feedback Nash equilibria, the cost function parameters must satisfy \eqref{eq:coupled_riccati} for the exactly recovered strategies, as established in Lemma~\ref{lem:iedentification_strategies}. We exploit this property to develop an inverse method for estimating the cost function parameters by reformulating \eqref{eq:coupled_riccati}, as described in the following.
\begin{lemma}\label{lem:reformulate_riccati}
    Let $\boldsymbol{\theta}_i \in \mathbb{R}^p$, with $p = 2n^2 + \sum_{j=1}^N m_j^2$, denote the vectorization of the matrices in \eqref{eq:cost_function} as
    \begin{equation}
    \begin{aligned}
         \boldsymbol{\theta}_i &=
        \big[
            \operatorname{vec}(\boldsymbol{Q}_i)^\top \,\,
            \operatorname{vec}(\boldsymbol{R}_{i1})^\top 
            \cdots\\
            &\operatorname{vec}(\boldsymbol{R}_{ii})^\top 
            \cdots \,\,
            \operatorname{vec}(\boldsymbol{R}_{iN})^\top\,\,
            \operatorname{vec}(\boldsymbol{Q}_{i,t_N})^\top
        \big]^\top,       
    \end{aligned}
    \end{equation}
    where $\operatorname{vec}(\boldsymbol{X})$ denotes the column-wise vectorization of a matrix $\boldsymbol{X}$.  
    Then, the matrices $\boldsymbol{Q}_i$, $\boldsymbol{R}_{ij}$, $\boldsymbol{Q}_{i,t_N}$ corresponding to $\boldsymbol{\theta}_i$ satisfy \eqref{eq:coupled_riccati} if and only if $\boldsymbol{\theta}_i$ satisfies
    \begin{equation}\label{eq:reformulation_riccati}
        \boldsymbol{M}_i(t)\boldsymbol{\theta}_i = \boldsymbol{0},\quad \forall t \in [t_0;t_N].
    \end{equation}
    The matrix $\boldsymbol{M}_i(t) \in \mathbb{R}^{m_i^2 \times p}$ in \eqref{eq:reformulation_riccati} is defined as
    \begin{equation}
    \begin{aligned}
        \boldsymbol{M}_i(t) &=
        \big[
            \boldsymbol{M}_{Q_i}(t) \,\,
            \boldsymbol{M}_{R_{i1}}(t)\,\, 
            \cdots\\
            &\boldsymbol{M}_{R_{ii}}(t)\,\, 
            \cdots\,\, 
            \boldsymbol{M}_{R_{iN}}(t)\,\,
            \boldsymbol{M}_{Q_{i,t_N}}
        \big],
    \end{aligned}
    \end{equation}
    with
    \begin{equation}\label{eq:definition_Ms}
        \begin{aligned}
            \boldsymbol{M}_{Q_i}(t) &= (t_N-t)(\boldsymbol{B}_i^\top \otimes \boldsymbol{B}_i^\top) \in \mathbb{R}^{m_i^2 \times n^2}, \\
            \boldsymbol{M}_{R_{ii}}(t) 
            &=\int_t^{t_N}\Big[\left(\left(\boldsymbol{B}_i^\top\boldsymbol{F}(s)^\top\boldsymbol{K}^*_i(s)^\top\right) \otimes \boldsymbol{I}_{m_i}\right)\\
            &+\left(\boldsymbol{I}_{m_i}\otimes \left(\boldsymbol{B}_i^\top\boldsymbol{F}(s)^\top \boldsymbol{K}^*_i(s)^\top\right)\right)\\
            &+\left(\boldsymbol{B}_i^\top\boldsymbol{K}^*_i(s)^\top\right) \otimes \left(\boldsymbol{B}_i^\top\boldsymbol{K}^*_i(s)^\top\right)\Big]ds\\
            &-\left(\left(\boldsymbol{B}_i^\top\boldsymbol{K}^*_i(t)^\top\right) \otimes \boldsymbol{I}_{m_i}\right)\in \mathbb{R}^{m_i^2 \times m_i^2}, \\
            \boldsymbol{M}_{R_{ij}}(t) &= \int_t^{t_N}\Big[\left(\boldsymbol{B}_i^\top\boldsymbol{K}^*_j(s)^\top\right)\\
            &\otimes \left(\boldsymbol{B}_i^\top\boldsymbol{K}^*_j(s)^\top\right)\Big]ds  \in \mathbb{R}^{m_i^2 \times m_j^2},\, j \in \mathcal{I} \setminus \{i\},\\
            \boldsymbol{M}_{Q_{i,t_N}} &= (\boldsymbol{B}_i^\top \otimes \boldsymbol{B}_i^\top) \in \mathbb{R}^{m_i^2 \times n^2},
            \end{aligned}
    \end{equation}
    where $\otimes$ is the Kronecker product.
\end{lemma}

\begin{pf}
    We rewrite \eqref{eq:coupled_riccati} as
    \begin{equation}
    \begin{aligned}
        \boldsymbol{0} &= \int_t^{t_N} \Big[\boldsymbol{R}_{ii} \boldsymbol{K}^*_i(s) \boldsymbol{F}(s)\boldsymbol{B}_i + \boldsymbol{B}_i^\top\boldsymbol{F}(s)^\top \boldsymbol{K}^*_i(s)^\top\boldsymbol{R}_{ii}\\
        &+\boldsymbol{B}_i^\top\boldsymbol{Q}_i\boldsymbol{B}_i
        + \sum_{j=1}^N \boldsymbol{B}_i^\top\boldsymbol{K}^*_j(s)^\top \boldsymbol{R}_{ij} \boldsymbol{K}^*_j(s)\boldsymbol{B}_i \Big] ds\\  
        &- \boldsymbol{R}_{ii} \boldsymbol{K}^*_i(t)\boldsymbol{B}_i + \boldsymbol{B}_i^\top\boldsymbol{Q}_{i,t_N}\boldsymbol{B}_i,
    \end{aligned}
    \end{equation}
    using \eqref{eq:relation_K_P}, the transversality condition, and integration. Next, applying the vectorization identity
    \begin{equation}
        \operatorname{vec}(\boldsymbol{X}\boldsymbol{Y}\boldsymbol{Z})
        = (\boldsymbol{Z}^\top \otimes \boldsymbol{X}) \operatorname{vec}(\boldsymbol{Y}),        
    \end{equation}
    we obtain
    \begin{equation}
    \begin{aligned}
        \boldsymbol{0}
        &= \int_t^{t_N} \Big[\left(\left(\boldsymbol{B}_i^\top\boldsymbol{F}(s)^\top\boldsymbol{K}^*_i(s)^\top\right) \otimes \boldsymbol{I}_{m_i}\right)\operatorname{vec}(\boldsymbol{R}_{ii}) \\
        &+ \left(\boldsymbol{I}_{m_i}\otimes \left(\boldsymbol{B}_i^\top\boldsymbol{F}(s)^\top \boldsymbol{K}^*_i(s)^\top\right)\right)\operatorname{vec}(\boldsymbol{R}_{ii}) \\
        &+(\boldsymbol{B}_i^\top \otimes \boldsymbol{B}_i^\top)\operatorname{vec}(\boldsymbol{Q}_i)\\
        &+ \sum_{j=1}^N \left(\left(\boldsymbol{B}_i^\top\boldsymbol{K}^*_j(s)^\top\right) \otimes \left(\boldsymbol{B}_i^\top\boldsymbol{K}^*_j(s)^\top\right)\right)\operatorname{vec}(\boldsymbol{R}_{ij}) \Big] ds\\
        &-\left(\left(\boldsymbol{B}_i^\top\boldsymbol{K}^*_i(t)^\top\right) \otimes \boldsymbol{I}_{m_i}\right)\operatorname{vec}(\boldsymbol{R}_{ii})\\
        &+(\boldsymbol{B}_i^\top \otimes \boldsymbol{B}_i^\top)\operatorname{vec}(\boldsymbol{Q}_{i,t_N}).
    \end{aligned}
    \end{equation}
    Rearranging the terms and substituting the definitions from \eqref{eq:definition_Ms} yields the reformulation in \eqref{eq:reformulation_riccati}.\qedsymbol
\end{pf}
In order to identify the unknown parameters $\boldsymbol{\theta}_i$ of \eqref{eq:cost_function}, we evaluate \eqref{eq:reformulation_riccati} at $K \in \mathbb{N}_+$ discrete time steps $t_k$. Stacking these $K$ systems of equations yields
\begin{equation}\label{eq:stacked_system_of_equations}
    \tilde{\boldsymbol{M}}_i \boldsymbol{\theta}_i :=
    \begin{bmatrix}
        \boldsymbol{M}_i(t_0) \\
        \boldsymbol{M}_i(t_1) \\
        \vdots \\
        \boldsymbol{M}_i(t_{K-1})
    \end{bmatrix}
    \boldsymbol{\theta}_i = \boldsymbol{0},
\end{equation}
where $\tilde{\boldsymbol{M}}_i \in \mathbb{R}^{K m_i^2 \times p}$.
To ensure that the game structure and \gls{GT} trajectories allow for unique identification of $\boldsymbol{\theta}_i$ up to a scalar factor, we introduce the following.
\begin{assum}\label{assum:rank}
    The matrix $\tilde{\boldsymbol{M}}_i$ satisfies $\operatorname{rank}(\tilde{\boldsymbol{M}}_i) = p - 1$.
\end{assum}
\begin{remark}
    A necessary condition for Assumption~\ref{assum:rank} to hold is that the input matrix satisfies $\operatorname{rank}(\boldsymbol{B}_i)=n$.
\end{remark}
\begin{lemma}\label{lem:theta_recovery}
    Let Assumption~\ref{assum:rank} hold. Then the parameter vector $\boldsymbol{\theta}_i$ satisfying \eqref{eq:coupled_riccati} can be recovered up to a scalar factor, using \eqref{eq:stacked_system_of_equations}.
\end{lemma}
\begin{pf}
    From Lemma~\ref{lem:necessary_sufficient}, the cost function parameters to be estimated must satisfy \eqref{eq:coupled_riccati}.  
    Lemma~\ref{lem:reformulate_riccati} establishes that \eqref{eq:reformulation_riccati} is an equivalent reformulation of \eqref{eq:coupled_riccati}.  
    By stacking $K$ such systems of equations as in \eqref{eq:stacked_system_of_equations} and ensuring that $\operatorname{rank}(\tilde{\boldsymbol{M}}_i) = p - 1$ via Assumption~\ref{assum:rank}, the null space of $\tilde{\boldsymbol{M}}_i$ is one-dimensional. This follows from the rank--nullity theorem (\cite{strang2012linear}, Section~2.4), 
since the dimension of the null space is
\begin{equation}
    \dim(\ker(\tilde{\boldsymbol{M}}_i)) 
    = p - \operatorname{rank}(\tilde{\boldsymbol{M}}_i) 
    = 1.
\end{equation}
    Hence, the solution to the homogeneous linear system \eqref{eq:stacked_system_of_equations} exists and is unique up to a scalar factor.  
    Consequently, the parameters $\boldsymbol{\theta}_i$ satisfying \eqref{eq:coupled_riccati} can be determined up to an arbitrary scaling.\qedsymbol
\end{pf}
\subsection{Identification of Noise Scaling Parameters}
Using an Euler--Maruyama discretization with step size $dt\in \mathbb{R}_+$, define the discrete-time residuals for demonstration $d$ as
\begin{equation}\label{eq:residuals}
    \Delta \boldsymbol{x}_k^{(d)} := \boldsymbol{x}_{k+1}^{(d)} - \boldsymbol{x}_k^{(d)} - \Big(\boldsymbol{A}\boldsymbol{x}_k^{(d)} + \sum_{i=1}^N \boldsymbol{B}_i \boldsymbol{u}_{i,k}^{(d)}\Big) dt 
    = \boldsymbol{L}\,\Delta\boldsymbol{\omega}_k^{(d)},
\end{equation}
where    $\Delta \boldsymbol{x}_k^{(d)} \sim \mathcal{N}(\boldsymbol{0},\,\boldsymbol{L}\boldsymbol{L}^\top dt)$. The log-likelihood $\mathcal{L}:\mathbb{R}^{n \times n} \to \mathbb{R}$ of all residuals across $D$ demonstrations and $K$ discrete time steps is then
\begin{equation}\label{eq:log_likelihood}
\begin{aligned}   
\mathcal{L}(\boldsymbol{L}\boldsymbol{L}^\top)
=&-\frac{1}{2} \sum_{d=1}^{D} \sum_{k=0}^{K-1} 
\Big(n \log(2\pi dt) + \log\det(\boldsymbol{L}\boldsymbol{L}^\top)\\
 &+ \frac{1}{dt}\,\Delta\boldsymbol{x}_k^{(d)\top} (\boldsymbol{L}\boldsymbol{L}^\top)^{-1} \Delta\boldsymbol{x}_k^{(d)}\Big).
\end{aligned}
\end{equation}

Maximizing \eqref{eq:log_likelihood} with respect to $(\boldsymbol{L}\boldsymbol{L}^\top)$ yields the maximum likelihood estimate
\begin{equation}\label{eq:L_hat_matrix}
    \hat{\boldsymbol{L}}\hat{\boldsymbol{L}}^\top
    = \frac{1}{D K dt} \sum_{d=1}^{D} \sum_{k=0}^{K-1} 
    \Delta\boldsymbol{x}_k^{(d)} \Delta\boldsymbol{x}_k^{(d)\top}.
\end{equation}
Since $\boldsymbol{L}$ is diagonal with $l_s>0,\, \forall s \in \{1,\dots,n\}$, we recover it from \eqref{eq:L_hat_matrix} as
\begin{equation}\label{eq:L_hat_diag}
    \hat{\boldsymbol{L}} = \operatorname{diag}\!\left(
        \sqrt{\hat{l}_1^2}, \dots,
        \sqrt{\hat{l}_n^2}
    \right).
\end{equation}

\begin{lemma}\label{lem:L_recovery}
Let the \gls{GT} trajectories be generated according to \eqref{eq:system} with noise scaling matrix $\boldsymbol{L}^*$. Then, as the total number of data points $DK \to \infty$, the maximum likelihood estimator \eqref{eq:L_hat_diag} converges to the true noise scaling matrix:
\begin{equation}\label{eq:L_true_recovery}
    \hat{\boldsymbol{L}} \to \boldsymbol{L}^* \quad \text{with probability 1}.
\end{equation}
\end{lemma}

\begin{pf}
The residuals $\Delta \boldsymbol{x}_k^{(d)}$ are independent and identically distributed Gaussian, with covariance $\boldsymbol{L}^* \boldsymbol{L}^{*\top} dt$. The total number of independent and identically distributed samples is $DK$, so by standard maximum likelihood consistency results (\cite{newey_large_nodate}, Theorem~2.5), the maximum likelihood estimate of $\hat{\boldsymbol{L}}\hat{\boldsymbol{L}}^\top$ converges almost surely to $\boldsymbol{L}^* \boldsymbol{L}^{*\top}$ as $DK \to \infty$. 
Since $\boldsymbol{L}^*$ is diagonal with positive entries, the positive square root of each diagonal element uniquely recovers $\boldsymbol{L}^*$, implying~\eqref{eq:L_true_recovery}. \qedsymbol
\end{pf}

\subsection{Inverse Stochastic Differential Game Framework}

We can solve Problem~\ref{prob:isdg} for the first time, combining the results from Lemma~\ref{lem:iedentification_strategies}, Lemma~\ref{lem:necessary_sufficient}, Lemma~\ref{lem:reformulate_riccati}, and Lemma~\ref{lem:theta_recovery} to estimate the cost function parameters $\boldsymbol{Q}_i, \boldsymbol{R}_{ij}, \boldsymbol{Q}_{i,t_N},\,\forall i \in \mathcal{I}$, with the results from Lemma~\ref{lem:L_recovery} to estimate the noise scaling matrix $\boldsymbol{L}$, both from \gls{GT} trajectories. The full \gls{ISDG} framework for \gls{LQG} differential games is summarized in Algorithm~\ref{algo:framework}.

\begin{algorithm}
\caption{Inverse \gls{LQG} Differential Game Framework}
\label{algo:framework}
\begin{algorithmic}[1]

\State \textbf{Input:} \parbox[t]{\dimexpr\linewidth-4em}{\gls{GT} trajectories $\{\boldsymbol{x}^{*(d)}(t_0 \to t_N)\}$ and $\{\boldsymbol{u}_i^{*(d)}(t_0 \to t_N)\},\, \forall i \in \mathcal{I}$, $d \in \{1,\dots,D\}$, and system matrices $\boldsymbol{A}$ and $\boldsymbol{B}_i,\, \forall i \in \mathcal{I}$.}

\State \textbf{Output:} \parbox[t]{\dimexpr\linewidth-5em}{Cost function parameters $\hat{\boldsymbol{Q}}_i$, $\hat{\boldsymbol{R}}_{ij}$, $\hat{\boldsymbol{Q}}_{i,t_N}$, $\forall i,j \in \mathcal{I}$, and noise scaling matrix $\hat{\boldsymbol{L}}$.}

\State {Identify feedback Nash strategies $\boldsymbol{K}^*(t)$ using \eqref{eq:exact_ls} from the \gls{GT} trajectories.}

\For{$i \in \mathcal{I}$}
    \State \parbox[t]{\dimexpr\linewidth-2em}{Construct the stacked matrix $\tilde{\boldsymbol{M}}_i$ for $K$ time steps as in \eqref{eq:stacked_system_of_equations}.}
    \State \parbox[t]{\dimexpr\linewidth-2em}{Estimate the cost function parameters $\hat{\boldsymbol{Q}}_i$, $\hat{\boldsymbol{R}}_{ij}$, $\hat{\boldsymbol{Q}}_{i,t_N}$, $\forall j \in \mathcal{I}$, by solving \eqref{eq:stacked_system_of_equations}.}
\EndFor

\State Compute the residuals $\Delta \boldsymbol{x}_k^{(d)}$ as in \eqref{eq:residuals}.
\State Estimate the covariance $\hat{\boldsymbol{L}}\hat{\boldsymbol{L}}^\top$ as in \eqref{eq:L_hat_matrix}.
\State Recover diagonal $\hat{\boldsymbol{L}}$ as in \eqref{eq:L_hat_diag}.
\State \Return $\hat{\boldsymbol{Q}}_i,\hat{\boldsymbol{R}}_{ij},\,\hat{\boldsymbol{Q}}_{i,t_N},\, \forall i,j \in \mathcal{I}$, and $\hat{\boldsymbol{L}}$.

\end{algorithmic}
\end{algorithm}

\begin{remark}
For the infinite-horizon case, the cost function parameters can be obtained via \cite{Ing:19} as part of the proposed \gls{ISDG} framework, with all other steps of Algorithm~\ref{algo:framework} unaffected.
\end{remark}

\section{Simulation Example}\label{sec:simulation}

To illustrate the proposed inverse \gls{LQG} differential game method, we consider a two-player example system
\begin{equation}\label{eq:eample_system}
    d\boldsymbol{x}
=
\left(\begin{bmatrix}
1 & -1 \\
1 & 0
\end{bmatrix}
\boldsymbol{x}
+
\begin{bmatrix}
1 & 0 \\
0 & 1
\end{bmatrix}
\boldsymbol{u}_1
+
\begin{bmatrix}
1 & 0 \\
0 & 1
\end{bmatrix}
\boldsymbol{u}_2\right)dt
+
\begin{bmatrix}
0.1 & 0 \\
0 & 0.2
\end{bmatrix}
d\boldsymbol{\omega}
\end{equation}
with initial state $\boldsymbol{x}_0=\left(2\quad-2\right)^\top$. The cost function matrices used to generate the \gls{GT} trajectories are 
$\boldsymbol{Q}^*_1 = \operatorname{diag}(1,1)$, 
$\boldsymbol{Q}^*_2 = \operatorname{diag}(1,10)$, 
$\boldsymbol{R}^*_{11} = \operatorname{diag}(1,1)$, 
$\boldsymbol{R}^*_{12} = \boldsymbol{0}$, 
$\boldsymbol{R}^*_{21} = \boldsymbol{0}$, 
$\boldsymbol{R}^*_{22} = \operatorname{diag}(1,2)$, and
$\boldsymbol{Q}_{1,t_N}^*=\boldsymbol{Q}_{2,t_N}^*=\boldsymbol{0}$. The \gls{GT} trajectories are generated by first solving the coupled Riccati differential equations to obtain the feedback Nash strategies, and then simulating the closed-loop system under these strategies to produce $D=20$ stochastic trajectories. Closed-loop stability of \eqref{eq:eample_system} was confirmed numerically. In our simulation study, the \gls{GT} trajectories serve as the input for the identification of the parameters $\hat{\boldsymbol{Q}}_i$, $\hat{\boldsymbol{R}}_{ij}$, $\hat{\boldsymbol{Q}}_{i,t_N}$, $\forall i,j \in \{1,2\}$, and $\hat{\boldsymbol{L}}$. 

\begin{figure}
    \centering
    \input{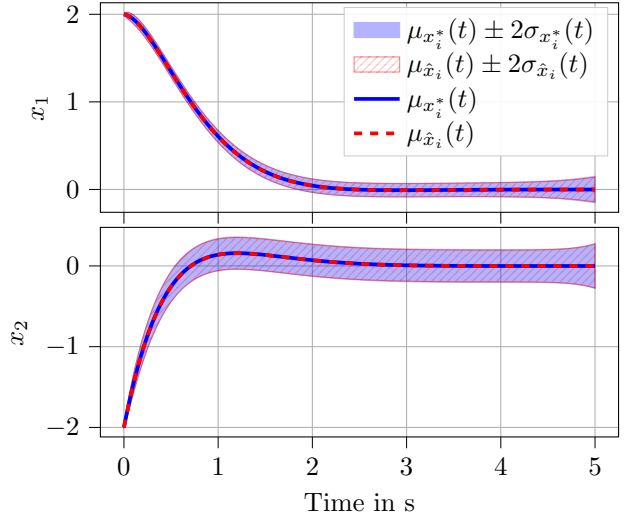}
    \caption{Estimated and GT state trajectories with mean and $\pm2\sigma$ confidence intervals for $D=20$ demonstrations and $K=20$ evaluated time steps, where $K=500$ is the full trajectory length.}
    \label{fig:state_mean_var}
\end{figure}

To evaluate the accuracy of the identification, we compute the forward solution using the estimated parameters and compare the resulting trajectories with the \gls{GT} trajectories. The comparison is based on the mean $\boldsymbol{\mu}_{\boldsymbol{x}} \in \mathbb{R}^2$ and variance $\boldsymbol{\sigma}^2_{\boldsymbol{x}} \in \mathbb{R}^2$ of the state trajectories, as well as the mean $\boldsymbol{\mu}_{\boldsymbol{u}_i} \in \mathbb{R}^{2}$ and variance $\boldsymbol{\sigma}^2_{\boldsymbol{u}_i} \in \mathbb{R}^{2}$ of the control input trajectories. The error metrics are defined as the maximum relative deviation between the \gls{GT} and estimated quantities, normalized by the maximum value of the corresponding \gls{GT} quantity. Specifically, for the state mean and control input mean we define
\begin{equation}\label{eq:error_x_mean}
      e^{\boldsymbol{\mu}_{\boldsymbol{x}}} = 
    \frac{\displaystyle \max_{t} \big\| \boldsymbol{\mu}_{\boldsymbol{x}^*}(t) 
    - \boldsymbol{\mu}_{\hat{\boldsymbol{x}}}(t) \big\|_\infty}
         {\displaystyle \max_{t} \big\| \boldsymbol{\mu}_{\boldsymbol{x}^*}(t) \big\|_\infty},  
\end{equation}
\begin{equation}\label{eq:error_u_mean}
    e^{\boldsymbol{\mu}_{\boldsymbol{u}}} 
    =\max_i \frac{\displaystyle \max_{t} 
    \big\| \boldsymbol{\mu}_{\boldsymbol{u}^*_{i}}(t) 
    - \boldsymbol{\mu}_{\hat{\boldsymbol{u}}_{i}}(t) \big\|_\infty}
         {\displaystyle \max_{t} 
    \big\| \boldsymbol{\mu}_{\boldsymbol{u}^*_{i}}(t) \big\|_\infty},
\end{equation}
where $\|\cdot\|_\infty$ denotes the maximum norm. For the state variance and control input variance, we define
\begin{equation}\label{error_x_var}
    e^{\boldsymbol{\sigma}^2_{\boldsymbol{x}}} =
    \frac{\displaystyle \max_{t} 
    \big\| \boldsymbol{\sigma}^2_{\boldsymbol{x}^*}(t)
    - \boldsymbol{\sigma}^2_{\hat{\boldsymbol{x}}}(t) \big\|_\infty}
         {\displaystyle \max_{t} 
    \big\| \boldsymbol{\sigma}^2_{\boldsymbol{x}^*}(t) \big\|_\infty},
\end{equation}
\begin{equation}\label{error_u_var}
    e^{\boldsymbol{\sigma}^2_{\boldsymbol{u}}} =
    \max_i\frac{\displaystyle \max_{t} 
    \big\| \boldsymbol{\sigma}^2_{\boldsymbol{u}_i^*}(t)
    - \boldsymbol{\sigma}^2_{\hat{\boldsymbol{u}}_i}(t) \big\|_\infty}
         {\displaystyle \max_{t} 
    \big\| \boldsymbol{\sigma}^2_{\boldsymbol{u}_i^*}(t) \big\|_\infty}.
\end{equation}
The total computation time $t_C \in \mathbb{R}$ represents the cumulative duration of all processing steps required to estimate the cost function and noise scaling parameters.
%for a given number of evaluated time steps $K$. 
To account for the stochasticity in the system, a batch-based study is conducted by repeating the forward simulation and identification procedure ten times. Finally, the averages over the ten repetitions of the four maximum errors \eqref{eq:error_x_mean}, \eqref{eq:error_u_mean}, \eqref{error_x_var} and \eqref{error_u_var} are reported. Figure~\ref{fig:state_mean_var} demonstrates for $K=20$ evaluated time steps that the estimated state mean and variance closely match the \gls{GT}, as indicated by the overlapping mean trajectories and $\pm2\sigma$ confidence intervals.

To further analyze the influence of the number of evaluated time steps, the batch-based study is repeated for $K \in \{20,\allowbreak 50,\allowbreak 100,\allowbreak500\}$, where $K = 500$ corresponds to the full trajectory length. The resulting averages of the maximum errors and total computation times are summarized in Table~\ref{tb:simulation_study}.

\begin{table}[hb]
\begin{center}
\caption{Averages of maximum errors and total computation time for different numbers of evaluated time steps $K$, $D=20$ demonstrations and ten repetitions.}
\label{tb:simulation_study}
\begin{tabular}{cccccc}
$K$ & $\overline{e}^{\boldsymbol{\mu}_{\boldsymbol{x}}}$ & $\overline{e}^{\boldsymbol{\mu}_{\boldsymbol{u}}}$ & $\overline{e}^{\boldsymbol{\sigma}^2_{\boldsymbol{x}}}$ & $\overline{e}^{\boldsymbol{\sigma}^2_{\boldsymbol{u}}}$ & $\overline{t}_C$ \\\hline
20  & 0.001885 & 0.060297 & 0.110194 & 0.137625 & 1.38$\,\mathrm{s}$ \\
50  & 0.000969 & 0.049900 & 0.084827 & 0.129632 & 3.31$\,\mathrm{s}$ \\
100 & 0.000304 & 0.003226 & 0.016994 & 0.011261 & 6.21$\,\mathrm{s}$ \\
500 & 0.000005 & 0.000047 & 0.016613 & 0.010992 & 16.80$\,\mathrm{s}$ \\
\end{tabular}
\end{center}
\end{table}

For $K=20$, the averages of the mean state error $\overline{e}^{\mu_{\boldsymbol{x}}}=1.89{\times}10^{-3}$ and the mean control input error $\overline{e}^{\mu_{\boldsymbol{u}}}=6.03{\times}10^{-2}$ are already small, indicating that even a short trajectory segment contains sufficient information to recover the cost function parameters with reasonable accuracy. The average errors of the state variance $\overline{e}^{\sigma^2_{\boldsymbol{x}}}=1.10{\times}10^{-1}$ and the control input variance $\overline{e}^{\sigma^2_{\boldsymbol{u}}}=1.37{\times}10^{-1}$ are notably higher, indicating that accurate estimation of the stochastic component requires more trajectory data. As $K$ increases, all error measures decrease systematically, with reductions by factors between 6.2 and 18.7 for $K=100$ compared to $K=20$. Beyond $K=100$, the averages of the variance errors saturate, suggesting that further improvements in variance estimation require additional trajectory demonstrations $(D>20)$ rather than longer single trajectories. The average of the total computation time $t_C$ increases with $K$, ranging from $\overline{t}_C=1.38\,\mathrm{s}$ for $K=20$ to $\overline{t}_C=16.8\,\mathrm{s}$ for $K=500$, reflecting the expected trade-off between accuracy and computational cost.

\section{Conclusion}\label{sec:conclusion}
We presented a method that solves the \gls{ISDG} problem for finite-horizon \gls{LQG} differential games. The proposed framework combines estimation of the feedback strategies via least squares estimation, identification of the cost function parameters through a novel reformulation of the coupled Riccati differential equations, and estimation of the noise scaling parameters via maximum likelihood estimation. Simulation results demonstrate that the approach accurately recovers both the cost function parameters of all players and the noise scaling parameters, yielding trajectories that closely match the observed \gls{GT} trajectories. Future work will focus on extending the proposed framework to differential games with further uncertainties.

\section*{DECLARATION OF GENERATIVE AI AND AI-ASSISTED TECHNOLOGIES IN THE WRITING PROCESS}
%\vspace{-3mm}
During the preparation of this work the authors used ChatGPT in order to perform language editing. After using this tool, the authors reviewed and edited the content as needed and take full responsibility for the content of the publication.

\bibliography{ifacconf}             % bib file to produce the bibliography

@book{Eng:05,
    author={J. Engwerda},
    title={LQ dynamic optimization and differential games},
    publisher={John Wiley \& Sons},
    year={2005},
	address={Chichester, England},
}

@book{BasOls:99,
    author={T. Başar and G.J. Olsder},
    title={Dynamic noncooperative game theory},
    publisher={SIAM},
    year={1999},
	address={Philadelphia, PA, USA},
}

@article{BucLi:08,
    author={R. Buckdahn and J. Li},
    title={Stochastic differential games and viscosity solutions of Hamilton–Jacobi–Bellman–Isaacs equations},
    journal={SIAM Journal on Control and Optimization},
    year={2008},
    volume={47},
    number={1},
    pages={444--475},
}

@article{Ham:98,
    author={S. Hamadène},
    title={Backward–forward SDE’s and stochastic differential games},
    journal={Stochastic Processes and their Applications},
    year={1998},
    volume={77},
    number={1},
    pages={1--15},
}

@book{Isa:65,
    author={R. Isaacs},
    title={Differential games: a mathematical theory with applications to warfare and pursuit, control and optimization},
    publisher={John Wiley \& Sons},
    year={1965},
	address={New York, NY, USA},
}

@inproceedings{Nat:10,
    author={S. Natarajan and G. Kunapuli and K. Judah and P. Tadepalli and K. Kersting and J. Shavlik},
    title={Multi-agent inverse reinforcement learning},
    booktitle={Ninth International Conference on Machine Learning and Applications},
    year={2010},
    pages={395--400},
}

@book{Mol:22,
    author={T.L. Molloy and J.I. Charaja and S. Hohmann and T. Perez},
    title={Inverse optimal control and inverse noncooperative dynamic game theory: a minimum-principle approach},
    publisher={Springer Nature},
    year={2022},
    address={Cham, Switzerland},
}

@article{Ing:19,
    author={J. Inga and E. Bischoff and T.L. Molloy and M. Flad and S. Hohmann},
    title={Solution sets for inverse non-cooperative linear-quadratic differential games},
    journal={IEEE Control Systems Letters},
    year={2019},
    volume={3},
    number={4},
    pages={871--876},
}

@article{Rot:17,
    author={S. Rothfuß and J. Inga and F. Köpf and M. Flad and S. Hohmann},
    title={Inverse optimal control for identification in non-cooperative differential games},
    journal={IFAC-PapersOnLine},
    year={2017},
    volume={50},
    number={1},
    pages={14909--14915},
}

@inproceedings{Che:24a,
    author={Z. Chen and L. Guo},
    title={An inverse problem for adaptive linear quadratic stochastic differential games},
    booktitle={63rd IEEE Conference on Decision and Control (CDC)},
    year={2024},
    pages={1838--1843},
}

@article{Meh:23,
    author={N. Mehr and M. Wang and M. Bhatt and M. Schwager},
    title={Maximum-entropy multi-agent dynamic games: forward and inverse solutions},
    journal={IEEE Transactions on Robotics},
    year={2023},
    volume={39},
    number={3},
    pages={1801--1815},
}

@article{Che:25,
    author={Z. Chen and L. Guo},
    title={Adaptive pursuit–evasion differential game with unknown cost functions},
    journal={Journal of Systems Science and Complexity},
    year={2025},
    volume={38},
    number={2},
    pages={533--546},
}

@article{TodJor:02,
    author={E. Todorov and M.I. Jordan},
    title={Optimal feedback control as a theory of motor coordination},
    journal={Nature Neuroscience},
    year={2002},
    volume={5},
    number={11},
    pages={1226--1235},
}

@inproceedings{Kil:24,
    author={S. Kille and P. Leibold and P. Karg and B. Varga and S. Hohmann},
    title={Human-variability-respecting optimal control for physical human–machine interaction},
    booktitle={33rd IEEE International Conference on Robot and Human Interactive Communication (RO-MAN)},
    year={2024},
    pages={1595--1602},
}

@inproceedings{Kar:24,
    author={P. Karg and M. Hess and B. Varga and S. Hohmann},
    title={Bi-level-based inverse stochastic optimal control},
    booktitle={European Control Conference (ECC)},
    year={2024},
    pages={537--544},
}

@article{MarCao:24,
    author={E. Martirosyan and M. Cao},
    title={Reinforcement learning for inverse linear–quadratic dynamic non-cooperative games},
    journal={Systems \& Control Letters},
    year={2024},
    volume={191},
}

@inproceedings{AbbNg:04,
    author={P. Abbeel and A.Y. Ng},
    title={Apprenticeship learning via inverse reinforcement learning},
    booktitle={Proceedings of the 21st International Conference on Machine Learning},
    year={2004},
    pages={1},
}

@inproceedings{Red:12,
    author={T.S. Reddy and V. Gopikrishna and G. Zaruba and M. Huber},
    title={Inverse reinforcement learning for decentralized non-cooperative multiagent systems},
    booktitle={IEEE International Conference on Systems, Man, and Cybernetics (SMC)},
    year={2012},
    pages={1930--1935},
}

@inproceedings{Yu:19,
    author={L. Yu and J. Song and S. Ermon},
    title={Multi-agent adversarial inverse reinforcement learning},
    booktitle={Proceedings of the 36th International Conference on Machine Learning},
    year={2019},
    pages={7194--7201},
}

@inproceedings{Zie:08,
    author={B.D. Ziebart and A. Maas and J.A. Bagnell and A.K. Dey},
    title={Maximum entropy inverse reinforcement learning},
    booktitle={Proceedings of the Twenty-Third AAAI Conference on Artificial Intelligence},
    year={2008},
    pages={1433--1438},
}

@article{MenZei:20,
    author={M. Menner and M.N. Zeilinger},
    title={Maximum likelihood methods for inverse learning of optimal controllers},
    journal={IFAC-PapersOnLine},
    year={2020},
    volume={53},
    number={2},
    pages={5266--5272},
}

@inproceedings{JeaMas:18,
    author={F. Jean and S. Maslovskaya},
    title={Inverse optimal control problem: the linear–quadratic case},
    booktitle={IEEE Conference on Decision and Control (CDC)},
    year={2018},
    pages={888--893},
}

@article{newey_large_nodate,
	title ={Large sample estimation and hypothesis testing},
	author={Whitney K. Newey and Daniel McFadden},
    journal={Handbook of econometrics},
    volume={4},
    pages={2111--2245},
    year={1994},
}

@inproceedings{Var:21,
  author={Varga, Balint and Inga, Jairo and Lemmer, Markus and Hohmann, S&#x00F6;ren},
  booktitle={2021 IEEE Conference on Control Technology and Applications (CCTA)}, 
  title={Ordinal Potential Differential Games to Model Human-Machine Interaction in Vehicle-Manipulators}, 
  year={2021},
  pages={728-734}
}

@book{strang2012linear,
  title={Linear algebra and its applications},
  author={Strang, Gilbert},
  year={2006},
  publisher = {Thomson, Brooks/Cole},
  address={Belmont, CA, USA}
}

@article{mylvaganam_differential_2017,
	author={Mylvaganam, Thulasi and Sassano, Mario and Astolfi, Alessandro},
    journal={IEEE Transactions on Automatic Control}, 
    title={A Differential Game Approach to Multi-agent Collision Avoidance}, 
    year={2017},
    volume={62},
    number={8},
    pages={4229-4235},
}

@article{gu_differential_2008,
	 author={Gu, Dongbing},
  journal={IEEE Transactions on Control Systems Technology}, 
  title={A Differential Game Approach to Formation Control}, 
  year={2008},
  volume={16},
  number={1},
  pages={85-93},
}

@article{LIAN2022110524,
title = {Inverse reinforcement learning for multi-player noncooperative apprentice games},
journal = {Automatica},
volume = {145},
pages = {110524},
year = {2022},
author = {Bosen Lian and Wenqian Xue and Frank L. Lewis and Tianyou Chai},
}
                                                     % with bibtex (preferred)

\end{document}